\documentclass[epj]{svjour}
\usepackage{graphics}
\begin{document}
\title{Uniaxially anisotropic antiferromagnets in a field on a square lattice}
\author{M. Holtschneider and W. Selke}
\institute{Institut f\"ur Theoretische Physik, Technische Hochschule
  RWTH Aachen, 52056 Aachen, Germany}
\date{Received: date / Revised version: date}
%
\abstract{
Classical uniaxially anisotropic Heisenberg and XY antiferromagnets
in a field along the easy axis on a square lattice
are analysed, applying ground state 
considerations and Monte Carlo techniques. The models are known to 
display antiferromagnetic and spin--flop phases. In the Heisenberg 
case, a single--ion anisotropy is added to the XXZ antiferromagnet, 
enhancing or competing with the uniaxial exchange anisotropy. Its 
effect on the stability of non--collinear structures of biconical 
type is studied. In the case of the anisotropic XY antiferromagnet, 
the transition region between the antiferromagnetic and spin--flop 
phases is found to be dominated by degenerate bidirectional 
fluctuations. The phase diagram is observed to resemble closely 
that of the XXZ antiferromagnet without single--ion anisotropy.
\PACS{
      {68.35.Rh}{Phase transitions and critical phenomena}   \and
      {75.10.Hk}{Classical spin models}   \and
      {05.10.Ln}{Monte Carlo method, statistical theory}
     } 
} 
\maketitle
\section{Introduction}
\label{intro}
Two--dimensional uniaxially anisotropic Heisenberg antiferromagnets in a
field along the easy axis have received a renewed interest in recent
years, experimentally as well as theoretically. On the experimental
side, especially layered cuprates exhibit interesting properties
due to an interplay of spin and
charge \cite{Mat1,Ammer,Pokro,Mat,Kroll,Rev1,Rev2,Schwing}.
Intriguing phase diagrams have been obtained for
other quasi two--dimensional antiferromagnets as
well, showing, typically, multicritical
behaviour \cite{Gaulin,Cowley,Chris,Pini}.

On the theoretical side, recent
studies \cite{Leidl1,Holt1,Zhou1,Leidl2,Holt2,Vic,Zhou2,Holt3} 
on the square lattice XXZ Heisenberg antiferromagnet
have substantially extended previous analyses
\cite{LanBin,Stein,Schmi,Costa} on this prototypical model.
The XXZ model is described by the Hamiltonian
\begin{equation}
  {\cal H}_{\mathrm{XXZ}} = J \sum\limits_{i,j} 
  \left[ \, \Delta (S_i^x S_j^x + S_i^y S_j^y) + S_i^z S_j^z \, \right] 
  \; - \; H \sum\limits_{i} S_i^z
\end{equation}

\noindent
where the sum runs over all pairs of neighbouring sites, $i$ and
$j$, of the lattice, $J (>0)$ is the coupling constant, and
$\Delta$ is the exchange anisotropy parameter, with $\Delta$=0 
corresponding to the Ising limit and $\Delta$= 1 to the isotropic
Heisenberg case. $H$ is the external field along 
the easy axis, the $z$--axis. $S_i^{\alpha}$, $\alpha$=$x,y$ and $z$,
are the three components of classical or quantum spins.

\begin{figure}
\resizebox{0.95\columnwidth}{!}{%
  \includegraphics{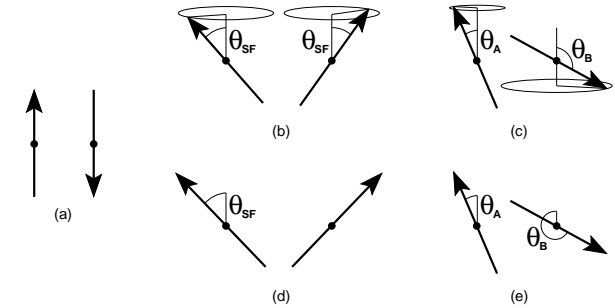}
}
\caption{Antiferromagnetic (a), spin--flop (b,d), and biconical (c)
as well as bidirectional (e) structures which may occur as ground states
in the XXZ antiferromagnet and variants, (a,b,c), or in the anisotropic XY
antiferromagnet (a,d,e). The tilt angles are defined with respect to
the easy axis, being the $z$--axis in the XXZ and the $x$--axis in
the XY case.}
\label{fig:1}
\end{figure}

For many years \cite{LanBin}, the model is known to display, at
low temperatures and small fields, a long--range ordered 
antiferromagnetic (AF) phase, and, at higher fields and 
low temperatures, a spin--flop (SF) phase with algebraically
decaying correlations. However, the multicritical point, where
the AF, SF and paramagnetic phases meet, had been subject to
controversial suggestions.

In early renormalisation
group calculations \cite{Nelson,Fish,KNFish} the 
XXZ Heisenberg model and its extensions to $n$--com\-po\-nent antiferromagnets
with uniaxial anisotropy have been investigated. Various possible
multicritical scenarios have been proposed, depending on the
number of spin components, $n$, and the dimension of
the lattice, $d$. The possible scenarios include a bicritical
point of $O(n)$ symmetry, a tetracritical point, and a 
critical end point. Actually, an $\epsilon$--expansion to low 
order, favours, for $n=3$, i.e. for the
XXZ antiferromagnet, the bicritical point. Certainly, the
bicritical point can not be realized in two dimensions at a 
non--zero temperature, $T >0$, because it would violate the
rigorously proven, well--known theorem of Mermin and Wagner \cite{MW}.

Recent Monte Carlo studies and ground--state considerations provide 
strong evidence for the multicritical point being a 
'hidden tetracritical point' at $T$=0 in the classical 
XXZ model \cite{Holt1,Zhou1,Holt2}. Indeed a narrow phase 
governed by 'biconical' \cite{KNFish} (BC) 
fluctuations separates the AF and SF phases at low temperatures. At that 
zero temperature special point, AF, SF, and BC structures have the same 
energy, leading to a high degeneracy \cite{Holt2}. Actually, the
importance of the non--collinear BC structures, see Fig. 1, for the
ground state and the phase diagram of the classical
XXZ antiferromagnet had been overlooked in previous work. On the
other hand, already a few decades ago, it had been noticed that
such BC structures may be stabilised by adding to the XXZ model
further anisotropy terms, like cubic terms, or 
longer--range interactions \cite{TM,LF,Wegner}. 

Extending significantly our very recent short
communication \cite{Holt3}, the aim of the present article is
twofold: Firstly, we shall study the effect of adding
a single--ion anisotropy to the XXZ model, which may either
enhance the uniaxial exchange anisotropy, $\Delta$, or it may
compete with it by being a planar anisotropy, depending
on the sign of the coupling strength of the single--ion
anisotropy. In both cases, the special point of high 
degeneracy (the hidden tetracritical point) does not
survive. Resulting phase diagrams will be determined
using Monte Carlo techniques. Secondly, we shall consider the
two--component, $n= 2$, variant of the XXZ
antiferromagnet. In that case, in principle, a bicritical
point, of $O(2)$ symmetry, would be allowed to occur
at $T>0$, being, in two dimensions, in the Kosterlitz--Thouless
universality class \cite{KT}. Note that now BC structures are
replaced by 'bidirectional' (BD) structures, see Fig. 1. Again, 
in addition to ground state considerations, the
phase diagram will be determined using Monte Carlo simulations.

The paper is organised as follows: In the next section, the
models will be introduced and ground state properties will
be discussed, emphasising the role of non--collinear, BC and
BD, structures. Phase diagrams and critical properties, as
obtained from large--scale simulations, will be presented in the 
then following section. A short summary concludes the article.

\section{Models and ground state properties}
In order to study the impact of biconical or bidirectional structures 
on phase diagrams of uniaxially anisotropic antiferromagnets
on a square lattice, we consider two different classical
models with spins of length one. Firstly, a single--ion
anisotropy is added to the XXZ model, eq. (1), so that the
Hamiltonian reads

\begin{equation}
{\cal H}_D = {\cal H}_{\mathrm{XXZ}} + D \sum\limits_{i} (S_i^z)^2
\end{equation}

\noindent
where the single--ion anisotropy may, depending on the sign of $D$,
enhance the uniaxial exchange anisotropy $\Delta$ ($0 \leq \Delta <
1$), when $D < 0$, or
it may introduce a competing planar anisotropy, $D > 0$. Secondly, the 
anisotropic XY antiferromagnet is studied, described by the 
Hamiltonian  
\begin{equation}
{\cal H}_{\mathrm{XY}} = J \sum\limits_{i,j} \left[ \, S_i^x S_j^x + \Delta S_i^y S_j^y \, \right] \; - \; H \sum\limits_{i} S_i^x
\end{equation}

\noindent
being the two--component, $n=2$, variant of the XXZ antiferromagnet,
where the $x$--axis is the easy axis.

The ground state configurations, at $T= 0$, are fixed by the spin
orientations on the two sublattices, $A$ and $B$, formed by
neighbouring sites of the square lattice. The configurations
may be determined in a straightforward way \cite{TM,LF,Holt4}.
 
For the XXZ model with single--ion anisotropy, eq.(2), the
$xy$--components of the spins order 
antiferromagnetically, having rotational symmetry. The
orientations of the spins on the two sublattices are then given
by their tilt angles, $\Theta_A$
and $\Theta_B$, with respect to the easy axis, the $z$--axis, see
Fig. 1. From minimisation of the energy, the actual values of the
tilt angles in the ground state configurations follow as
a function of the anisotropy parameters $\Delta$ and $D$ as well
as the field $H$. For calculational convenience, one may substitute
$\Theta_A$ and $\Theta_B$ by combinations of the $z$--components
of the sublattice
magnetisations (per sublattice site), $S_A^z= \cos \Theta_A$
and $S_B^z=  \cos \Theta_B$.

\begin{figure}
\resizebox{0.95\columnwidth}{!}{%
  \includegraphics{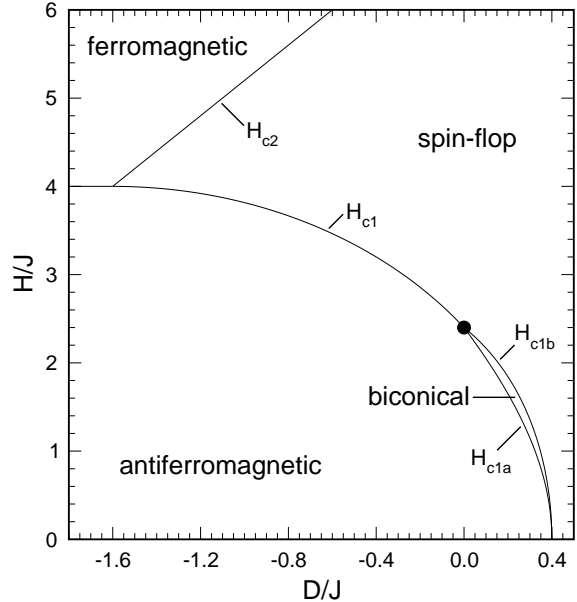}
}
\caption{Ground state phase diagram for $\Delta = 0.8$. The filled
  circle denotes the highly degenerate point at $D=0$ and $H=2.4 J$.}
\label{fig:2}
\end{figure}

\noindent
For negative couplings $D<0$, only AF, SF, and
and ferromagnetic (F) ground states occur, see Fig. 2 (setting
$\Delta$= 0.8, as before \cite{Holt1,Zhou1,Holt2,Holt3,LanBin}). As
depicted in the figure, AF
ground states are stable for low fields, $H < H_{c1}$, while F
ground states are obtained for high fields, $H > H_{c2}$. The two
critical fields are given by, for $D > -2\Delta J$, 

\begin{equation}
 H_{c1}= 2 \sqrt{ (2J)^2 - (2J\Delta + D)^2 }
\end{equation}

\noindent
and

\begin{equation}
 H_{c2}= 4J (1+\Delta) + 2D
\end{equation}

\noindent
For intermediate fields $H$, $H_{c1} < H < H_{c2}$, the ground states comprise
SF configurations, but they are squeezed out for a more negative single--ion
term, $D < -2 \Delta J$, fostering the uniaxial
alignment of the spins, see Fig. 2.

Note that for $D < 0$, in the limit $\Delta = 1$, when the
uniaxial anisotropy of
the Hamiltonian is solely due to the single--ion term, no
biconical structures exist
as ground states. In contrast, when the uniaxial anisotropy is
solely due to the exchange anisotropy, i.e. in the XXZ
antiferromagnet, $D= 0$, BC structures are ground states at the critical
field $H_{c1} = 4J \sqrt{1- \Delta^2}$, see
eq. (4) \cite{Holt2,Holt3}. At this highly degenerate point the
BC structures take on tilt angles interrelated by \cite{Holt2}

\begin{equation}
 \Theta_B = \arccos \left( \frac{ \sqrt{1-\Delta^2} \; - \; \cos\Theta_A }{ 1 \; - \; \sqrt{1-\Delta^2} \cos\Theta_A } \right)
\end{equation}

\noindent
interpolating continuously between the AF and SF configurations, with
the tilt angle $\Theta_A$ ranging from 0 to $\pi$. One may 
calculate ground state values of various quantities  at the degenerate
point, assuming that each degenerate configuration with the
interrelated tilt angles occurs with the same probability and
taking into account the rotational invariance of the spin components
in the $xy$--plane. 

\begin{figure}
\resizebox{0.95\columnwidth}{!}{%
  \includegraphics{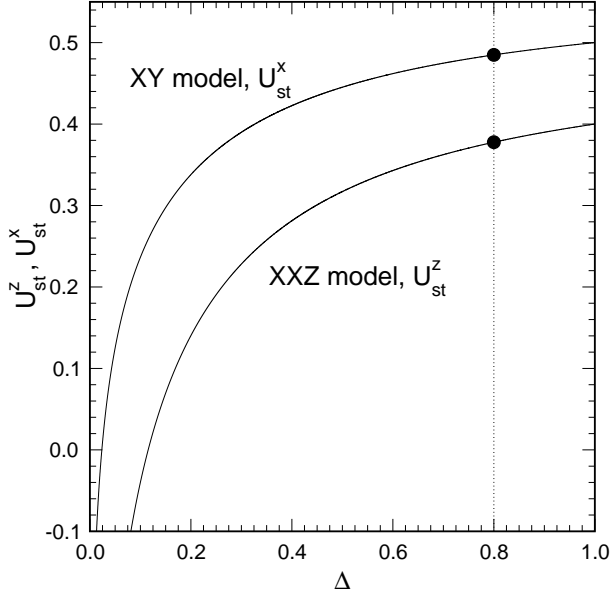}
}
\caption{Binder cumulants $U_{st}^{x,z}$ of the longitudinal staggered
 magnetisations of the XXZ and anisotropic XY antiferromagnets 
 at the highly degenerate point $T= 0$ and $H_{c1}$.}
\label{fig:3}
\end{figure}

\noindent
As an example we mention the Binder cumulant \cite{Binder}

\begin{equation}
  U_{st}^z = 1 - <(M_{st}^z)^4>/(3 <(M_{st}^z)^2>)
\end{equation}

\noindent
where the longitudinal staggered magnetisation $M_{st}^z$ is defined
by $M_{st}^z= (S_A^z-S_B^z)/2$. The cumulant
at the highly degenerate point $(T=0, H=H_{c1})$ as a function of
$\Delta$ is shown in
Fig. 3. For instance, in the much studied
case $\Delta =0.8$, one obtains $U_{st}^z= 0.3777...$ \cite{Holt4}. In
simulating small systems at low temperatures on approach to the degenerate
point for various values of $\Delta$, we confirmed the (numerically)
exact result for the cumulant.

An interesting quantity for characterising the BC structures is the
probability of finding the tilt angle $\Theta$, $p(\Theta)$. At the 
degenerate point, one gets \cite{Holt4}

\begin{equation}
  p(\Theta) = \frac{\alpha\;\sin\Theta }{1 - \alpha\, \cos\Theta} \; \left[ \, \ln \left( \, \frac{1+\alpha}{1-\alpha} \, \right) \, \right]^{-1}
\end{equation}

\noindent
where $\alpha= \sqrt{1- \Delta^2}/2$. Obviously, there is no
full $O(3)$ symmetry, as one may expect for a bicritical point of
Heisenberg type. One may reproduce that form of $p(\Theta)$ by
simulating, again, small systems at low temperatures on
approach to the highly degenerate point.

We now consider the case of a single--ion
anisotropy favouring a planar
anisotropy, $D > 0$, competing with the uniaxial exchange
anisotropy $\Delta$. Now, biconical structures may become
ground states in a finite, non--zero range
of fields, $H_{c1a} < H < H_{c1b}$, as depicted in Fig. 2. The
critical field between the AF and BC ground states, $H_{c1a}$, is
given by

\begin{equation}
  H_{c1a} = 2 \sqrt{ (2J-D)^2 - (2J\Delta)^2 }
\end{equation}

\noindent
and the upper critical field, $H_{c1b}$, separating the BC
and SF ground states, is given by

\begin{equation}
  H_{c1b} = 4 \left[ \, 4 J^2 - (2J\Delta + D)^2 \, \right] / H_{c1a}
\end{equation}

\noindent
At $D = 2-2\Delta$, the two critical fields approach zero, and, at
larger planar anisotropies, there exists no AF ground state. 

Of course, the degeneracy in the BC structures occurring
at the special point $(T=0,H=H_{c1})$ of the XXZ model is
now lifted, with the tilt angles, $\Theta_A$, $\Theta_B$, changing
now continuously with the field $H$, starting with the AF structure
at $H_{c1a}$ and ending with the SF structure at $H_{c1b}$. The
magnetisations on the $A$ and $B$ sublattices in this biconical 
region are interrelated by \cite{Holt4}

\begin{equation}
  S_A^z + S_B^z = \sqrt{1-\frac{(2 J \Delta)^2}{(2J+D)^2}} \; (1+S_A^zS_B^z)
\end{equation}

\noindent
The expression transforms into eq.(6) for vanishing single--ion
anisotropy, $D=0$.

The ground state properties of the anisotropic XY
antiferromagnet, eq.(3), are closely related to those of the XXZ
model, eq. (1). There
is also a high degeneracy at the critical field separating the
AF and SF ground states, $H_{c1}/J= 4 \sqrt{1-\Delta^2}$, in
this case due to bidirectional structures. As depicted in
Fig. 1, the BD configurations are characterised by tilt
angles, $\Theta_A$ and $\Theta_B$, which
are again defined with respect to the easy axis, being now
the $x$--axis. Note that here the tilt angles may vary
from 0 to $2 \pi$. At the highly degenerate point, $\Theta_A$ and $\Theta_B$
are interrelated analogously to eq. (6).

At $T=0$ and $H= H_{c1}$, various quantities of interest may be 
calculated, assuming again that each configuration with interrelated
tilt angles, being an AF, a BD or a SF state, occurs with the same
probability. In contrast to the XXZ case, there is now no
planar rotational invariance to be taken into account. For example,
the dependence of the Binder cumulant of the longitudinal staggered
magnetisation, $U_{st}^x$, on the exchange anisotropy $\Delta$ is
included in Fig. 3. The probability $p(\Theta)$ for encountering the
tilt angle $\Theta$ is found to be

\begin{equation} 
 p(\Theta) = \frac{1}{1 - \alpha\, \cos\Theta} \; 
  \frac{\sqrt{1-\alpha^2}}{2 \pi} 
\end{equation}

\noindent
where $\alpha= \sqrt{1-\Delta^2}/2$. Analogously to the XXZ
case, there is no full $O(2)$ symmetry. Again, these ground state results have
been checked in simulations, as for the XXZ case.

\section{Phase diagrams}
To study the effect of the presence of BC or BD
structures on the phase diagrams of
uniaxially anisotropic antiferromagnets on a square lattice, we
shall consider
the XXZ Heisenberg antiferromagnet with an additional single--ion
anisotropy, eq. (2), as well as the anisotropic XY 
antiferromagnet, eq. (3). In all
cases the exchange anisotropy anisotropy is set $\Delta= 0.8$, as
before.

Large--scale Monte Carlo simulations have been performed, studying
lattices with up to $L^2= 240^2$ spins, with runs of typically
up to, for larger lattices, $10^8$ Monte Carlo steps per
spin, averaging over several realizations to estimate
standard deviations for the computed quantities. These
quantities include the specific heat, sublattice, staggered, and
total magnetisations, longitudinal (relative to the direction
of the applied field) and transverse staggered susceptibilities,
Binder cumulants and related histograms. To monitor BC and BD
fluctuations and structures, we recorded probability functions
of the tilt angles, such as the
probability $p_2(\Theta_A,\Theta_B)$ for finding the
two angles, $\Theta_A$ and $\Theta_B$, at
neighbouring sites and the probability $p(\Theta)$ for encountering the tilt
angle $\Theta$. The
probabilities may be defined 'locally' by taking into account
the orientations of individual spins, or 'globally' by computing
average spin orientations on the two sublattices. Obviously, the
two definitions coincide at zero temperature.

To determine transition temperatures, the finite--size dependence 
of various quantities has been recorded, with reasonable 
extrapolations to the thermodynamic limit (the exact finite--size
dependence is not always known). The estimates we obtained
agreed within the error bars shown in the phase
diagrams.

\subsection{XXZ antiferromagnets with single--ion anisotropy}
Let us consider first the case of a negative single--ion
anisotropy, $D <0$, enhancing the exchange anisotropy. In this case, there are
no ground states of BC type. In Fig. 4, a 
typical phase diagram is depicted, where $D/J= -0.2$.  

\begin{figure}
\resizebox{0.95\columnwidth}{!}{%
  \includegraphics{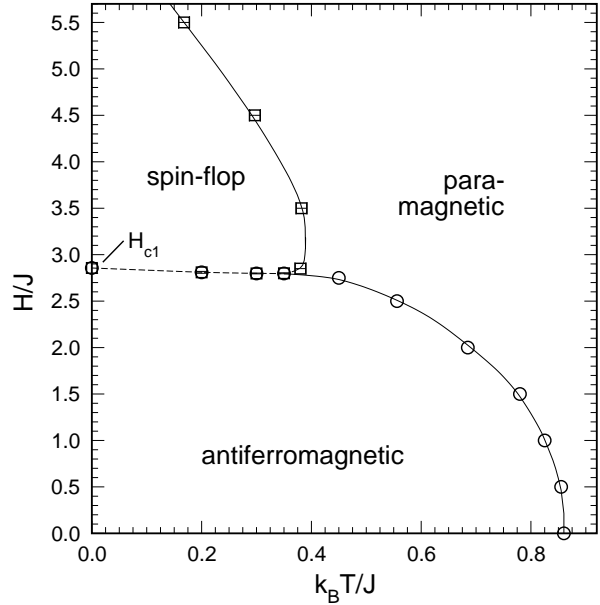}
}
\caption{Phase diagram of the XXZ antiferromagnet with a single--ion
anisotropy, $\Delta= 0.8$ and $D/J= -0.2$.}
\label{fig:4}
\end{figure}

\noindent
At low temperatures, we observe a transition of first order
separating the AF and SF phases. Evidence for that kind of
phase transition has been presented before \cite{Holt3}. For
instance, the maximum of the longitudinal staggered susceptibility is
seen to grow with system size, $L$, like $L^a$, and $a$ approaches
very closely 2 already for quite small sizes, $L \ge 10$. That
behaviour is characteristic for a first--order transition with
a rather small correlation length at the transition. 

Further evidence for a first--order transition may be
inferred from the
probability $p(\Theta)$ for finding the tilt angle $\Theta$. Close
to the transition, $p(\Theta)$ shows
more and more pronounced
local maxima simultaneously at the values of $\Theta$ characterising
the AF phase as well as the SF
phase, when increasing the system size. Note that at least for small system
sizes, biconical fluctuations are also observed in the transition
region between the AF and SF phases, but the relevant effect 
seems to be the coexistence phenomenon. 

To identify the nature of the triple point, where the AF, SF, and paramagnetic
phases meet, a very fine resolution, in temperature and
field, is needed near that
point. This aspect, requiring, presumably, huge computational
efforts, is beyond the scope of the present study.

\begin{figure}
\resizebox{0.95\columnwidth}{!}{%
  \includegraphics{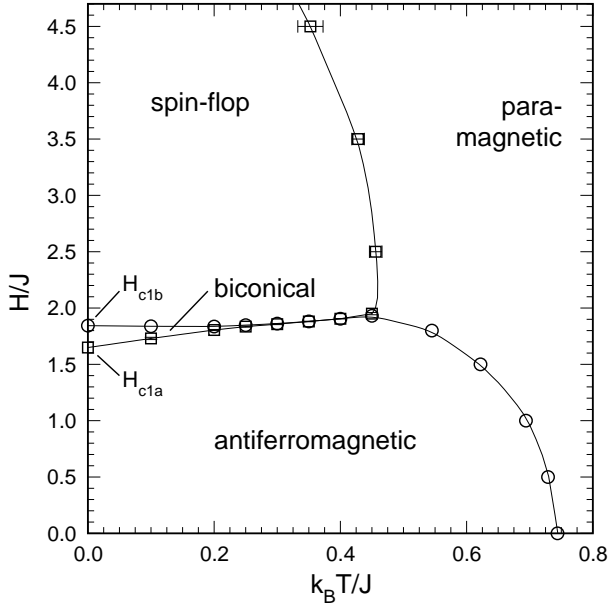}
}
\caption{Phase diagram of the XXZ antiferromagnet with a competing
single--ion anisotropy, $\Delta= 0.8$ and $D/J= 0.2$.}
\label{fig:5}
\end{figure}

\noindent
When applying a planar single--ion anisotropy, $D> 0$, competing 
with the uniaxial exchange anisotropy $\Delta$, a very different phase
diagram results. An example is shown in Fig. 5
for $D/J= 0.2$. In this case, biconical structures occur as ground
states in a non--zero range of fields, bordered by $H_{c1a}$
and $H_{c1b}$, see eqs. (9) and (10). At $T>0$, they are expected
to give rise to an ordered BC phase, in which
the ordered AF and SF phases
coexist \cite{TM,LF}. Actually, here in two dimensions, the algebraic
order of the SF phase, is found to vanish at the boundary of
the AF and BC phases, at $H_{c1a}$, while $M_{st}^z$, the order parameter
of the AF phase, vanishes at the higher critical field
$H_{c1b}$, separating the BC and SF phases, compare to Fig. 5.

\begin{figure}
\resizebox{0.95\columnwidth}{!}{%
  \includegraphics{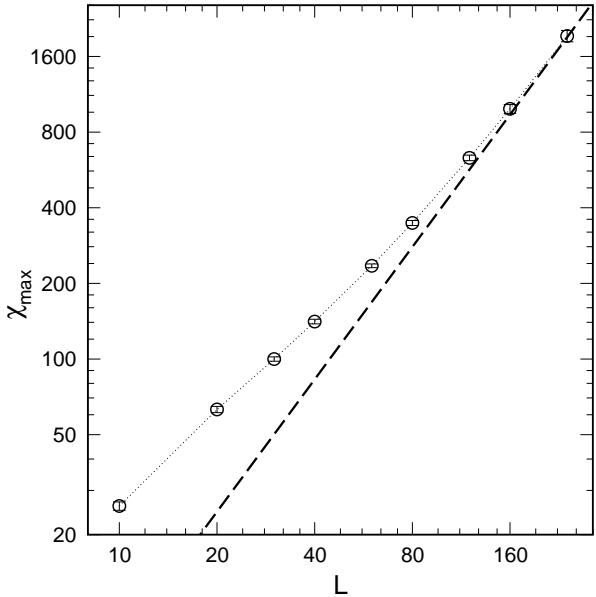}
}
\caption{Critical exponent for staggered susceptibility $\chi$ for
the XXZ antiferromagnet with a competing single--ion
anisotropy, at $k_BT/J=0.2$, see Fig. 5.}
\label{fig:6}
\end{figure}

\noindent
Based on renormalisation group calculations
\cite{Bruce,Aharony,Mukamel,Domany}, the transition
between the BC and SF phases may be argued to be in the Ising universality
class, while the transition between the BC and AF phases 
is expected to be in the XY universality class, being the
Kosterlitz--Thouless universality class \cite{KT} in two
dimensions. 

This description is in accordance with our simulational data. For
instance, we monitored the size--dependence of the maximum
of the longitudinal staggered susceptibility, $\chi_{max}(L)$, being
located close to $H_{c1b}$, see Fig. 6 for $k_BT/J=0.2$. From
the doubly logarithmic plot shown in that figure, one observes
that the effective exponent $a$, defined
by $\chi_{max} \propto L^a$, seems to approach, indeed, the asymptotic
Ising value of 7/4 for rather large system
sizes, $L \ge 120$. Thence, significant corrections to scaling
play an important 
role. In turn, at the boundary line between the BC and AF phases
the algebraic order
in the transverse staggered magnetisation, which holds in the
BC phase, gets lost. The finite-size dependence of that
magnetisation, for $L \geq 40$,
agrees with the transition belonging to the
Kosterlitz--Thouless universality class, where the order parameter
vanishes at the transition in the form of a power--law with an
exponent $\eta= 1/4$.

\begin{figure}
\resizebox{0.95\columnwidth}{!}{%
  \includegraphics{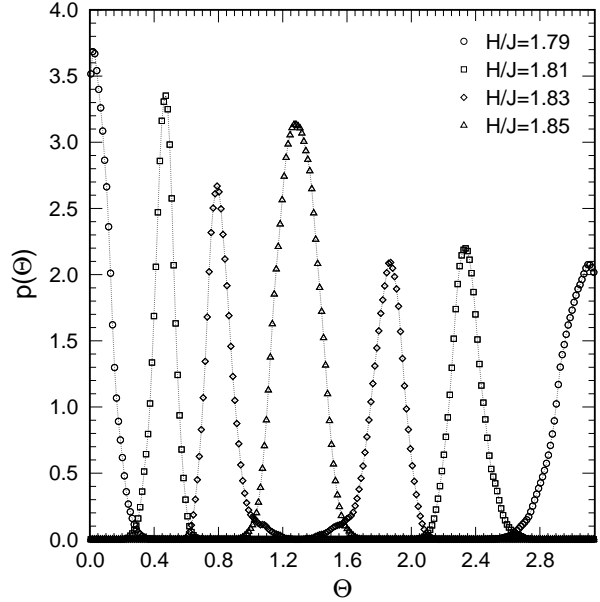}
}
\caption{Histograms for the probability of the
tilt angle $p( \Theta)$ for the XXZ antiferromagnet with a competing
single--ion anisotropy, $D/J= 0.2$, at $k_BT/J= 0.2$, at the fields given
in the inset. Lattices with $80^2$ spins are simulated. Note that
the area under the curve has been normalised to be one.}
\label{fig:7}
\end{figure}

\noindent
In the BC phase the dominant interrelated tilt angles are 
changing continuously, at fixed low temperature, with
the field. In fact, this behaviour is displayed by
the probability function $p(\Theta)$, as illustrated in Fig. 7
for the global (sublattice) spin orientations. By increasing
the field, at $k_BT/J=0.2$, the peak positions correspond first to the AF
structure, shifting gradually towards each other, reflecting BC
structures, and finally collapsing
in one peak characterising the SF phase.
  
As seen in Fig. 5, the extent of the BC phase shrinks with
increasing temperature. Eventually, the BC phase may terminate
at a tetracritical point \cite{LF,Bruce,Aharony,Mukamel,Domany}, where 
the AF, SF, BC, and paramagnetic phases meet. Because the phase
boundaries seem to meet there with common tangents, we give here
only a rough estimate for the case depicted in Fig. 5 ($D/J= 0.2$ and
$\Delta= 0.8$), $k_BT_{tetra}/J= 0.35 \pm 0.05$. A more precise location
and an analysis of its critical properties seem to
require enormous computational efforts.

\subsection{Anisotropic XY antiferromagnet}

Finally, let us consider the anisotropic
XY model, setting the exchange anisotropy $\Delta$= 0.8. Its
phase diagram is depicted in Fig.8.

\begin{figure}
\resizebox{0.95\columnwidth}{!}{%
  \includegraphics{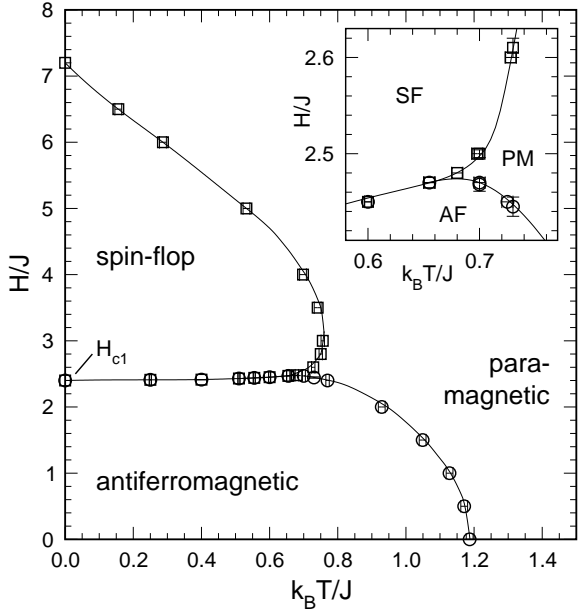}
}
\caption{Phase diagram of the anisotropic XY model with $\Delta=0.8$.}
\label{fig:8}
\end{figure}

The topology of the phase diagram looks like in the XXZ
case \cite{Holt1,Zhou1,Holt2}. The AF and SF boundary lines approach
each other very closely near the maximum of the AF phase boundary in the
$(T,H)$--plane. Accordingly, at low temperatures, there
seems to be either a direct transition between the AF and
SF phases, or two separate transitions with an extremely
narrow intervening phase may occur. At zero temperature
and $H= H_{c1}(= 2.4J)$, the highly degenerate ground state comprises
SF, AF, and bidirectional structures.

Away from that intriguing transition region, see Fig. 8, one expects the
transitions between the paramagnetic and the AF as well as
the SF phases to be
in the Ising universality class, because in the SF phase of the XY
antiferromagnet, there is just one ordering
component, the $y$--component. This consideration
is confirmed by the Monte Carlo data for the specific
heat (where the peak at the
AF phase boundary gets rather weak on approach to the transition
region) and for the staggered
susceptibilities. The quantities exhibit critical behaviour of
Ising--type, as
follows from the corresponding effective exponents describing size
dependences of the various peak heights.

\begin{figure}
\resizebox{0.95\columnwidth}{!}{%
  \includegraphics{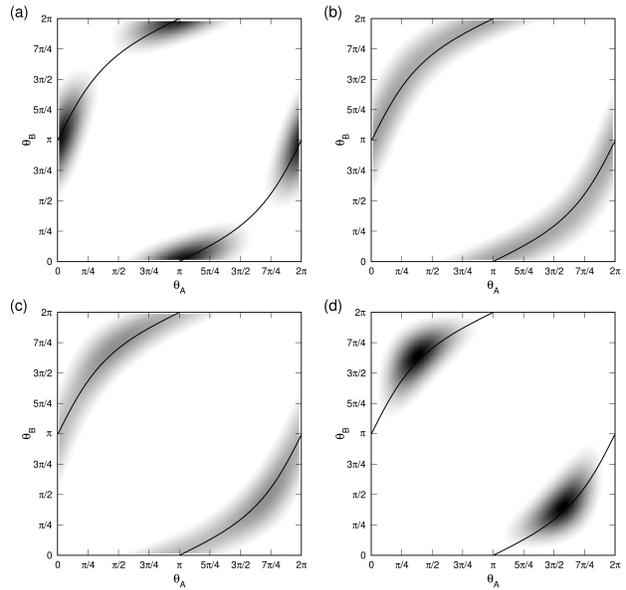}
}
\caption{Probability $p_2(\theta_A,\theta_B)$ for
    the anisotropic XY antiferromagnet with $\Delta$= 0.8 for a system
    with $100 \times 100$ lattice sites near the transition region between
    the AF and SF phases at $k_BT/J= 0.4$ and
    $H/J=$ (a) 2.40, (b) 2.4136, (c) 2.4138, and (d) 2.43.
    $p_2(\theta_A,\theta_B)$ is proportional to the
    grayscale. The superimposed solid line depicts the relation between
    the two tilt angles in the ground state.}
\label{fig:9}
\end{figure}

In the transition region of the AF and SF phases, BD
fluctuations dominate,
as one may conveniently infer from the (local) probability distribution
$p_2(\Theta_A,\Theta_B)$ for finding the two tilt angles 
at neighbouring sites, i.e. for the two different
sublattices. Typical results are 
depicted in Fig. 9, showing the behaviour of $p_2$ in a
grayscale representation, at fixed temperature, $k_BT/J= 0.4$ and
different fields, somewhat below and above the, possibly, two closeby
transitions as well as in their immediate vicinity. Indeed, at the
low field, see Fig. 9(a), one
finds a distribution corresponding to the AF phase, with peaks at 
$\Theta_{A/B}= 0, \pi$. At the high field, Fig. 9(d), there is a
distribution typical for the SF phase, with peaks at the spin--flop
tilt angle. Note that the local maxima in $p_2$ follow
closely the line describing the dependence of the two tilt
angles in the degenerate ground state, eq. (6), albeit
the probability of these BD structures is low compared to
those of the AF and SF, resp., structures, see Figs. 9(a) and 9(d). In
other words, thermal
fluctuations driving the system away from the AF or SF
structures are rather weak and, predominantly, of BD type.

In the immediate vicinity of the transitions, see
Figs. 9(b) and 9(c), the BD fluctuations and structures
clearly dominate. Now, all those
degenerate bidirectional structures occur simultaneously
with (almost) equal probability, i.e. along
the line of maxima $p_2$ is (almost) constant. Actually, it is
also interesting to monitor  
the time evolution of Monte Carlo configurations and of
the probability $p(\Theta)$ in those configurations. One notices
that at a given time of the simulation a 
concrete combination of interrelated tilt angles 
prevails. As Monte Carlo time evolves, other combinations in accordance
with the ground state degeneracy prevail, leading to the
behaviour depicted in Fig. 9. Obviously, this is is marked contrast to
the situation in the ordered biconical phase at fixed field
and temperature. There, after equilibration, just one combination
of tilt angles seems to dominate during the entire simulation, see Fig. 7.

Note that in the transition region at higher temperatures, analysing
systems of fixed size, say, $L= 100$ (see Fig. 9), the
various BD structures tend
to occur simultaneously in a single MC
configuration. Thence, for each configuration, $p(\Theta)$ has a shape 
quite similar to that in the degenerate ground state. This  
observation may be explained by a smaller correlation length. Typically,
when increasing the system size, the region of
(almost) constant values of $p_2$ along the line of local maxima
shrinks somewhat. Certainly, further systematic studies of the time
scales and temperature as well as  size dependences related
to the BD structures would be desirable.

Critical phenomena in the transition region
between the AF and SF phases
have been studied also by analysing effective exponents. We did that for
the staggered susceptibilities and the specific heat. Results
are compatible with Ising--type criticality, but rather large corrections
to scaling had to be presumed. For example, at $H/J=2.44$
and $0.54 < k_BT/J < 0.57$, see Fig. 8, the effective critical
exponents for describing the size--dependences of the peak height for
the staggered susceptibilities are about
1.8 to 1.85, largely independent of system size. The supposedly
rather strong corrections to scaling may be due to very large correlation
lengths in that region, and the asymptotics  may be reached
only for very large systems.

In any event, the simulational data seem to
suggest for the anisotropic XY antiferromagnet the existence of an extremely
narrow, disordered phase, intervening between the AF
and SF phases, like in the XXZ case \cite{Holt1,Zhou1}. That
intermediate phase is dominated by all the, in the
ground state completely degenerate, bidirectional fluctuations.

\section{Summary}
\label{sec:1}

In this article, we studied two variants of the XXZ antiferromagnet
in a field along the easy axis on a square
lattice, by, firstly, adding a single--ion
anisotropy, and by, secondly, reducing the number of spin components to two,
yielding the anisotropic XY antiferromagnet. Large--scale
Monte Carlo simulations have been performed, augmented by
ground state calculations.  

Adding a single--ion anisotropy has a drastic impact both on
ground state properties and the phase diagram. Biconical
structures, leading to a highly degenerate ground state in
the XXZ antiferromagnet, are either
suppressed, when the single--ion anisotropy fosters the
uniaxial exchange anisotropy, or their degeneracy will be lifted
by stabilising them successively with changing field, when the
single--ion anisotropy introduces a planar anisotropy. In the
former case, we observe, at low temperatures a direct first--order transition
between the AF to SF phases, while in the other case, an ordered
biconical phase emerges, separating the AF and SF phases. The 
situation in the XXZ case without single--ion anisotropy, where
a narrow disordered phase with biconical fluctuations separates
the AF and SF phases, interpolates in between these two 
scenarios with single--ion anisotropies of different sign.

When the uniaxiality of the antiferromagnet is solely due to
a single--ion anisotropy, i.e. when the exchange couplings are
isotropic, no biconical structures occur as ground states.    

Ground state properties and the phase diagram of the
anisotropic XY antiferromagnet are observed to resemble rather closely those
of the XXZ antiferromagnet. There is a highly degenerate
ground state, at which non--collinear
structures of bidirectional type become stable. These degenerate
bidirectional structures prevail at low temperatures in the transition region
between the AF and SF phases, leading, presumably, to a very narrow
intervening disordered phase.

We conclude that ground state properties and
phase diagrams of classical uniaxially anisotropic
antiferromagnets in two dimensions depend crucially on
the form of the anisotropy terms, supporting or suppressing
non--collinear structures of biconical or bidirectional type. 

We should like to thank especially A. Aharony, G. Bannasch,
K. Binder, D. P. Landau,  A. Pelissetto, and E. Vicari for useful
correspondence, information, remarks, and discussions. We gratefully
acknowledge financial support by the Deutsche Forschungsgemeinschaft under
grant SE324/4.


\begin{thebibliography}{}
\bibitem{Mat1} M.\ Matsuda, K.\ M.\ Kojima, Y.\ J.\ Uemura, J.\ L.\ Zarestky,
  K.\ Nakajima, K.\ Kakurai, T.\ Yokoo, S.\ M.\ Shapiro, and G.\ Shirane, Phys.\ Rev.\ B \textbf{57}, 11467 (1998).
\bibitem{Ammer} U.\ Ammerahl, B.\ B{\"u}chner, C.\ Kerpen,
  R.\ Gross, and A.\ Revcolevschi,
  Phys.\ Rev.\ B\ \textbf{62}, R3592 (2000).
\bibitem{Pokro} W.\ Selke, V.\ L.\ Pokrovsky, B.\ B{\"u}chner, and
  T.\ Kroll, Eur.\ Phys.\ J.\ B\ \textbf{30}, 83 (2002); M.\
  Holtschneider and W.\ Selke, Phys. Rev. E \textbf{68}, 026120 (2003).
\bibitem{Mat} M.\ Matsuda, K.\ Kakurai, J.\ E.\ Lorenzo, L.\ P.\ Regnault,
  A.\ Hiess, and G.\ Shirane, Phys.\ Rev.\ B \textbf{68}, 060406(R) (2003).
\bibitem{Kroll} T.\ Kroll, R.\ Klingeler, J.\ Geck, B.\ B{\"u}chner,
  W.\ Selke, M.\ H{\"u}cker, and A.\ Gukasov, J.\ Magn.\ Magn.\ Mat.\ \textbf{290}, 306 (2005).
\bibitem{Rev1} M.\ Uehara, N.\ Motoyama, M.\ Matsuda, H.\ Eisaki, and J.\ Akimitsu,
  in: A.\ V.\ Narlikar (Ed.), {\it Frontiers in Magnetic Materials},
  Springer (2005).
\bibitem{Rev2} T.\ Vuleti{\'c}, B.\ Korin-Hamzi{\'c}, T.\ Ivek, S.\ Tomi{\'c},
  B.\ Gorshunov, M.\ Dressel, and J.\ Akimitsu,
  Phys.\ Rep.\ \textbf{428}, 169 (2006).
\bibitem{Schwing} U.\ Schwingenschl{\"o}gl and C.\ Schuster, 
  Europhys.\ Lett.\ \textbf{79}, 27003 (2007); 
  Phys. Rev. Lett.\ \textbf{99}, 237206 (2007).
\bibitem{Gaulin} B.\ D.\ Gaulin, T.\ E.\ Mason, M.\ F.\ Collins, and J.\ Z.\ Larese, 
  Phys.\ Rev.\ Lett.\ \textbf{62}, 1380 (1989).
\bibitem{Cowley} R.\ A.\ Cowley, A.\ Aharony, R.\ J.\ Birgeneau, R.\ A.\ Pelcovits, G.\ Shirane, and T.\ R.\ Thurston,
  Z.\ Phys.\ B\ \textbf{93}, 5 (1993).
\bibitem{Chris} R.\ J.\ Christianson,
  R.\ L.\ Leheny, R.\ J.\ Birgeneau, and R.\ W.\ Erwin, Phys.\ Rev.\ B\ \textbf{63}, 140401(R) (2001).
\bibitem{Pini} M.\ G.\ Pini, A.\ Rettori, P.\ Betti, J.\ S.\ Jiang,
  Y.\ Ji, S.\ G.\ E.\ te Velthuis, G.\ P.\ Felcher, and S.\ D.\ Bader,
  J.\ Phys.: Condens. Matter\ \textbf{19}, 136001 (2007).
\bibitem{Leidl1} R.\ Leidl and W.\ Selke,
  Phys. Rev. B \textbf{70}, 174425 (2004); Phys. Rev B \textbf{69}, 056401 (2004).
\bibitem{Holt1} M.\ Holtschneider, W.\ Selke, and R.\ Leidl,
  Phys.\ Rev.\ B\ \textbf{72}, 064443 (2005).
\bibitem{Zhou1} C.\ Zhou, D.\ P.\ Landau, and T.\ C.\ Schulthess,
  Phys.\ Rev.\ B\ \textbf{74}, 064407 (2006).
\bibitem{Leidl2} R.\ Leidl, R.\ Klingeler, B.\ B{\"u}chner, M.\ Holtschneider, and W.\ Selke,
  Phys.\ Rev.\ B\ \textbf{73}, 224415 (2006).
\bibitem{Holt2} M.\ Holtschneider, S.\ Wessel, and W.\ Selke,
  Phys.\ Rev.\ B\ \textbf{75}, 224417 (2007).
\bibitem{Vic} A.\ Pelissetto and E.\ Vicari,
  Phys.\ Rev.\ B\ \textbf{76}, 024436 (2007).
\bibitem{Zhou2} C.\ Zhou, D.\ P.\ Landau, and T.\ C.\ Schulthess,
  Phys.\ Rev.\ B\ \textbf{76}, 024433 (2007).
\bibitem{Holt3} M.\ Holtschneider and W.\ Selke,
  Phys.\ Rev.\ B\ \textbf{76}, 220405(R) (2007).
\bibitem{LanBin} K.\ Binder and D.\ P.\ Landau,
  Phys.\ Rev.\ B\ \textbf{13}, 1140 (1976);
  D.\ P.\ Landau and K.\ Binder,
  Phys.\ Rev.\ B\ \textbf{24}, 1391 (1981).
\bibitem{Stein} R.\ van\ de\ Kamp, M.\ Steiner and H.\ Tietze-Jaensch,
  Physica B\ \textbf{241-243}, 570 (1997).
\bibitem{Schmi} G.\ Schmid, S.\ Todo, M.\ Troyer, and A.\ Dorneich,
  Phys.\ Rev.\ Lett. \textbf{88}, 167208 (2002).
\bibitem{Costa} R.\ Costa and W.\ Pires,
  J.\ Mag.\ Mag.\ Mat. \textbf{262}, 316 (2003).
\bibitem{Nelson} D.\ R.\ Nelson, J.\ M.\ Kosterlitz, and M.\ E.\ Fisher,
  Phys. Rev. Lett. \textbf{33}, 813 (1974).
\bibitem{Fish} M.\ E.\ Fisher and D.\ R.\ Nelson,
  Phys. Rev. Lett. \textbf{32}, 1350 (1974).
\bibitem{KNFish} J.\ M.\ Kosterlitz, M.\ E.\ Fisher, and D.\ R.\ Nelson,
  Phys. Rev. B \textbf{13}, 412 (1976).
\bibitem{MW} N.\ D.\ Mermin and H.\ Wagner,
  Phys. Rev. Lett. \textbf{17}, 1133 (1966).
\bibitem{TM} H.\ Matsuda and T.\ Tsuneto,
  Prog.\ Theoret.\ Phys.\ Suppl.\ \textbf{46}, 411 (1970).
\bibitem{LF} K.-S.\ Liu and M.\ E.\ Fisher,
  J.\ Low.\ Temp.\ Phys.\ \textbf{10}, 655 (1973).
\bibitem{Wegner} F.\ Wegner, Solid State Commun.\ \textbf{12}, 785 (1973).
\bibitem{KT} J.\ M.\ Kosterlitz and D.\ J.\ Thouless,
  J.\ Phys.\ C\ \textbf{6}, 1181 (1973).
\bibitem{Holt4} M.\ Holtschneider, 
  PhD thesis, RWTH Aachen (2007).
\bibitem{Binder} K.\ Binder,
  Z.\ Phys.\ B\ \textbf{43}, 119 (1981); 
  Phys.\ Rev.\ Lett.\ \textbf{47}, 693 (1981).
\bibitem{Bruce} A.\ D.\ Bruce and A.\ Aharony,
  Phys.\ Rev.\ B \textbf{11}, 478 (1975).
\bibitem{Aharony} A.\ Aharony,
  J.\ Stat.\ Phys.\ \textbf{110}, 659 (2003).
\bibitem{Mukamel} D.\ Mukamel,
  Phys.\ Rev.\ B\ \textbf{14}, 1303 (1976).
\bibitem{Domany} E.\ Domany and M.\ E.\ Fisher,
  Phys. Rev. B \textbf{15}, 3510 (1977).
\end{thebibliography}
\end{document}